\documentstyle[12pt]{article}
\begin{document}

\textheight=8.5in
\textwidth=6in
\hfuzz=6pt
\let\fef=\ref
\def\ref#1{(\fef{#1})}
\def\SUN{{\sl SU(N)}}
\def\Trace{{\rm Tr\,}}

\title{Equivalence of Many-Gluon Green Functions in Duffin-Kemmer-Petieu and 
Klein-Gordon-Fock Statistical Quantum Field Theories}

\author{B.M.Pimentel\thanks{Instituto de Fisica, Universidade Estadual Paulista, 
S\~ao Paulo, Brazil, E-mail: pimentel@ift.unesp.br.} 
{}\ and V.Ya.Fainberg\thanks{P.N.Lebedev Institute of 
Physics, Moscow, Russia, E-mail: fainberg@lpi.ru.}}
\date{}

\maketitle

\begin{abstract}
We prove the equivalence of many-gluon Green functions in Duffin-Kemmer-Petieu (DKP)
and Klein-Gordon-Fock (KGF) statistical quantum field theories.  The proof is based on
the functional integral formulation for the statistical generating functional in a 
finite-temperature quantum field theory.  As an illustration, we calculate one-loop
polarization operators in both theories and show that their expressions indeed coincide.
\end{abstract}


\section{Introduction}

This work is a straightforward generalization of the articles [1]--[3] which established the 
equivalence of many-photon Green functions in DKP and KGF statistical quantum field theories.

In Section~2 we present a general proof of equivalence using the functional integral 
method in statistical quantum field theory. From the physical viewpoint, our result 
is understandable qualitatively: in non-zero temperature conditions gluons do not
become massive (and thus do not acquire any chemical potential), so that their 
intrinsic properties remain the same.  To illustrate this, in Section~3 we calculate
one-loop polarization operators in both theories and show that these operators 
actually coincide.  Section~4 contains conclusive remarks.

\section{Coincidence of Many-Gluon Green Functions in DKP and KGF Theories for 
Finite Temperatures}

To construct a generating functional $Z(J,\bar J,J_\mu)$ for the Green functions 
(GF) in statistical theory, we should perform transition to Euclidean space and then
limit the integration area along $x_4$: $0\le x_4\le\beta$, where $\beta=1/T$, $T$ 
is the temperature, and $J,\bar J,J_\mu$ are the external currents.  For simplicity, 
from now on we restrict ourselves to the case of {\it fundamental\/} representation 
of the \SUN\ group (see [4], [5]).

In DKP theory, the functional integral describing interaction between the gluon
field $A^a_\mu$ and charged particles of spin-0 and mass $m$ has the following form
(in the $\alpha$-gauge):
\begin{eqnarray}\label{1}
&&\kern-10pt Z_{\rm DKP}(J^i,\bar J^j,J_\mu) = Z_0\int_\beta
DA_\mu^a D\psi^i D\bar\psi^i \exp\left\{\left[ 
-{1\over4}F_{\mu\nu}^a F_{\mu\nu}^a - {1\over 2\alpha}(\partial_\mu A_\mu^a)^2
\right. \right.\nonumber\\
&&\kern-10pt{} - \bar{c}^i(\partial_\mu D_\mu^{ij}c^j) +
\bar\psi^i\left(i\beta_\mu 
D_\mu^{ij}-m\delta^{ij}\right)\psi^j+\bar{J}^i\psi^i %\nonumber\\ &&\qquad\qquad{} 
+ J^i\bar\psi^i + A_\mu^a J_\mu^a\biggr]d^4x\biggr\}.
\end{eqnarray}
Here $a = 1,2,\ldots N^2-1$ is the group index; $i,j=1,2,\ldots N_j$, 
\begin{eqnarray}\label{2}
&& D_\mu^{ij} = \delta^{ij}\partial_\mu - ig\left(A_\mu^a t^a\right)^{ij} 
\nonumber\\
&& [t_a,t_b]_- = if_{abc}t_c,
\end{eqnarray} 
where $f_{abc}$ are the \SUN\ group structure constants.

The Euclidean $\beta_\mu$-matrices in Eq.\ref{1} are chosen as in [3] (see 
Eq.(4)); the fields $c^i,\bar c^j$ are the Faddeev-Popov ghosts [6].
As for the functional integrations in this formula, they are understood as
\begin{equation}\label{3}
\int_\beta DA_\mu^a = \int\prod_{0\le x_4\le\beta}\
\prod_{-\infty\le x_i\le+\infty}\
\prod_{a=1}^{N^2-1}\ \prod_\mu\ dA_\mu^a.
\end{equation}

To prove the coincidence of many-gluon Green functions, let us integrate out the
$\psi^i$ and $\bar\psi^j$ fields in Eq.\ref{1}.  We get:

\begin{eqnarray}\label{4}
&&\kern-10pt Z_{\rm DKP}(J^i,\bar J^j,J_\mu) = Z_0\int_\beta
DA_\mu^a \exp\left\{-\int_\beta d^4x\left[ 
{1\over4}F_{\mu\nu}^a F_{\mu\nu}^a + {1\over 2\alpha}(\partial_\mu A_\mu^a)^2
\right. \right.\nonumber\\
&&\qquad\qquad{} - \bar{c}^i(\partial_\mu D_\mu^{ij}c^j) 
+ J_\mu^a A_\mu^a  + \Trace\ln S^{ii}(x,x,A_\mu^a)\biggr] \nonumber\\ 
&&\qquad\qquad\qquad{} - \int_\beta d^4x\,d^4y\, \bar J^i (x) S^{ij}(x,y,A^a)
J^j(y)\biggr\},
\end{eqnarray}
where
\begin{equation}\label{5}
S^{ij}(x,y,A) = \left(\beta_\mu D_\mu^{ij}-m\delta^{ij}\right)^{-1}\delta(x-y)
\end{equation}
is the Green function for the DKP-particle in the external Yang-Mills field $A_\mu^a$.
The term $\Trace\ln S(x,x,A)$ gives rise to all vacuum diagrams in perturbation 
theory.

On the other hand,
\begin{eqnarray}\label{6}
&&\exp\Trace\ln S^{ij} (x,x,A^a) = \det S^{ij} (x,y,A^a) \nonumber \\
&&\quad{} = \int_\beta D\psi^i D\bar\psi^j \exp
\left\{
-\int_\beta d^4x\, \bar\psi^i(x) 
\left(\beta_\mu D_\mu^{ij}+\delta^{ij}m\right)
\psi^j(x)
\right\},
\end{eqnarray}
where $\psi^i(x)$ is the column vector
\begin{equation}\label{7}
\psi^i(x) = 
\left(
\begin{array}{r}
\phi^i(x) \\
\partial_4 \phi^i(x) \\
\partial_1 \phi^i(x) \\
\partial_2 \phi^i(x) \\
\partial_3 \phi^i(x) 
\end{array}
\right),
\end{equation}
and thus we can rewrite Eq.\ref{6} in the component form as:
\begin{eqnarray}\label{8}
&&\exp\Trace\ln S^{ij} (x,x,A^a) = \det S^{ij} (x,y,A^a) \nonumber \\
&&\qquad{} = \int_\beta \prod_{\mu=1}^4
D_\mu \phi^{*i} D_\mu \phi^{j} D\phi^{*i} D\phi^{j}
\exp\biggl\{-\int_\beta d^4x\,\biggl[
\phi^{*i} D^{ij}_\mu \phi_\mu^j 
\nonumber \\ &&\qquad\qquad{} 
+ \phi^{*i}_\mu D_\mu^{ij}\phi^j + m
\left(\phi^i\phi^{*i}+\phi_\mu^{*i}\phi_\mu^i\right)
\biggr]\biggr\}.
\end{eqnarray}
After integrating over $\phi_\mu^{*i}$ and $\phi_\mu^i$ we obtain
\begin{eqnarray}\label{9}
&& \det S^{ij} (x,y,A^a)\equiv \det G^{ij}(x,y,A) = \exp\Trace\ln G^{ii}(x,x,A)\nonumber \\
&&{} = \int_\beta
D\phi^i D\phi^{*j}\exp\left\{
+{1\over m}\int_\beta d^4x \phi^{*i}\left((D^2_\mu)^{ij}-m^2\delta^{ij}\right)\phi^j
\right\},
\end{eqnarray}
where
\begin{equation}\label{10}
G^{ij}(x,y,A^a) = \left(-(D^2_\mu)^{ij}+m^2\delta^{ij}\right)\delta^4(x-y)
\end{equation}
is the Green function of the KGF equation in the external field $A^a_\mu(x)$.

It follows from Eqs.\ref{8}--\ref{10} that many-gluon Green functions coincide 
in DKP and KGF theories.  This completes the proof of equivalence of many-gluon
Green functions in these theories.

\section{Polarization Operator in One-Loop Approximation}
In order to prove the equality of one-loop polarization operators in DKP and KGF statistical
theories it is sufficient to consider the loops formed by scalar massive particles,
since all other one-loop diagrams coincide in these theories.

The one-loop polarization operator in DKP theory has the following form:
\begin{equation}\label{11}
\Pi_{\mu\nu}^{\rm DKP}(k) = {g^2\over (2\pi)^2\beta}\Trace\int
d{\bf p}\beta_\mu(t^a)^{ij} S^{jk}(p+k)\beta_\nu(t^a)^{kl}S^{li}(p),
\end{equation}
where 
\begin{equation}\label{12}
S^{jk}(p) = \delta^{jk}(i\hat p - m)^{-1}.
\end{equation}

One easily checks that
\begin{eqnarray}
&& \label{13} S^{jk} = \delta^{jk}(i\hat p-m)^{-1} = 
-{\delta^{ik}\over m}\left(\frac{i\hat p(i\hat p+m)}{p^2+m^2}+1\right), \\
&& \hat p = \beta_\mu p_\mu,\ p^2 = p_4^2+{\bf p}^2,\ p_4 = {2\pi n\over\beta},\
-\infty < n < + \infty, \nonumber \\
&& \label{14} (i\hat p-m)^{ij}S^{jk}(\hat p) = \delta^{ik}.
\end{eqnarray}

Using Eq.\ref{11}--\ref{14} we obtain the polarization operator in DKP theory
(in $g^2$-approximation):
\begin{eqnarray}\label{15}
&& \Pi_{\mu\nu}^{\rm DKP}(k) = {g^2\over (2\pi)^2\beta}
(t_a)^{ij}(t_a)^{ji}\sum_{p_4}\int d{\bf p}\biggl(
\frac{(2p+k)_\mu(2p+k)_\nu}{(p^2+m^2)((p+k)^2+m^2)} \nonumber\\
&&\qquad\qquad\qquad\qquad{}-\frac{\delta_{\mu\nu}}{p^2+m^2}
-\frac{\delta_{\mu\nu}}{(p+k)^2+m^2} \biggr).
\end{eqnarray}

In KGF theory, the one-loop polarization operator equals to:\footnote{The expressions
\ref{15} and \ref{16} coincide with Eqs.(14) and~(10) in [3] up to the substitution
$e^2 \to g^2(t_a)^{ij}(t_a)^{ji}$.}
\begin{equation}\label{16}
\Pi_{\mu\nu}^{\rm KGF}(k) = {g^2(t^a)^{ij}(t^a)^{ji}\over (2\pi)^3\beta}
\sum_{p_4} \int d{\bf p} \left(
\frac{(2p+k)_\mu(2p+k)_\nu}{(p^2+m^2)((p+k)^2+m^2)} - \frac{2\delta^{\mu\nu}}{p^2+m^2}
\right).
\end{equation}
The term proportional to $\delta_{\mu\nu}$ in Eq.\ref{16} plays an important role
in the proof of transversality of $\Pi_{\mu\nu}$ (i.e., $k_\mu\Pi_{\mu\nu}=0$). This 
term appears because of the term $\sim (A^a_\mu(x))^2$ entering $(D^2_{\mu})^{ij}$ 
in Eq.\ref{10}.

After the substitution $(p+k)\to p$ in the term $\delta_{\mu\nu}((p+k)^2+m^2)^{-1}$ 
of Eq.\ref{15} it changes into $\delta_{\mu\nu}(p^2+m^2)^{-1}$, so that the right-hand
sides of Eq.\ref{15} and Eq.\ref{16} coincide, and become formally gauge-invariant.
This coincidence of $\Pi_{\mu\nu}^{\rm DKP}$ and $\Pi_{\mu\nu}^{\rm KGF}$ in one-loop
approximation confirms the general proof of equivalence presented in Section~2.

In relativistic quantum field theory, the $\Pi_{\mu\nu}(k)$ tensor has the form:
\begin{equation}\label{17}
\Pi_{\mu\nu} = \left(k_\mu k_\nu - k^2\delta_{\mu\nu}\right)\Pi(k^2).
\end{equation}

In quantum statistics, $\Pi_{\mu\nu}$ depends on the two vectors: $k_\mu$ and 
$u_\mu$, the latter being the unit vector of the media velocity. Thus, the most general
expression reads (see [7], page 75)
\begin{eqnarray}\label{18}
&& \Pi_{\mu\nu} =  \biggl(\delta_{\mu\nu}-{k_\mu k_\nu\over k^2}\biggr)A_1 +
\biggl(
u_\mu u_\nu - \frac{k_\mu u_\nu(ku)}{k^2} \nonumber \\ &&\qquad\qquad{}
- \frac{k_\nu u_\mu (ku)}{k^2} 
 + \frac{k_\mu k_\nu(ku)^2}{k^2}
\biggr)A_2  \equiv \Phi_{\mu\nu}^1 A_1 + \Phi_{\mu\nu}^2 A_2.
\end{eqnarray}
Introducing the notation (in any approximation)
\begin{eqnarray}\label{19}
&& a_1 \equiv \Pi_{\mu\mu} = 3A_1+\lambda A_2 \nonumber \\
&& a_2 = u_\mu \Pi_{\mu\nu}u_\nu = \lambda(A_1 + \lambda A_2),\quad 
\lambda=1-{(ku)^2\over k^2},
\end{eqnarray}
we have
\begin{equation}\label{20}
A_1 = {1\over 2} \left(a_1 - {1\over\lambda}a_2\right),\quad
A_2 = {1\over 2\lambda} \left(-a_1 + {3\over\lambda}a_2\right).
\end{equation}

If the system is at rest\footnote{One can show that the representation \ref{18} is, strictly
speaking, valid only if the system is at rest: ${\bf u}=0, u_4=1$ (see [7], Chapter 11, Section 7).}
\begin{equation}\label{21}
\lambda = 1 - {k^2_4\over k^2}
\end{equation}
and
\begin{equation}\label{22}
a_2 = \left(1-{k_4^2\over k^2}\right)A_1 +
\left(1-{k_4^2\over k^2}\right)^2A_2.
\end{equation}

To proceed further, it is convenient to represent $a_1$ and $a_2$ in the form
$$ a_i = a_i^R + a_i^\beta,\quad i=1,2.$$

Here the terms $a_i^R$ do not depend on $\beta$ and must be renormalized; 
$a_i^\beta$ depend on $\beta$ and must vanish when $\beta\to\infty$:
\begin{equation}\label{23}
\lim_{\beta\to\infty}a_i^\beta = 0.
\end{equation}

Now Eq.\ref{18} may be rewritten in the following form:
\begin{eqnarray}\label{24}
\Pi_{\mu\nu}={1\over2}\Phi^1_{\mu\nu}
\bigl(a_1^R-{1\over\lambda}a_2^R+a_1^\beta-{1\over\lambda}a_2^\beta\bigr) + 
{1\over2\lambda}\Phi^2_{\mu\nu}
\bigl(-a_1^R+{3\over\lambda}a_2^R-a_1^\beta+{3\over\lambda}a_2^\beta\bigr).
\end{eqnarray}

The terms $\sim\Phi_{\mu\nu}^2$ should vanish when $\beta\to\infty$. 
Correspondingly, in this limit we get the $\Pi_{\mu\nu}$ tensor of Euclidean quantum
field theory.  Since $a_1^R$ and $a_2^R$ do not depend on $\beta$, we obtain after 
renormalization:
\begin{equation}\label{25}
a_2^R = {\lambda\over 3}a_1^R.
\end{equation}
Thus
\begin{equation}\label{26}
\lim_{\beta\to\infty}\Pi_{\mu\nu} = {1\over3}\Phi_{\mu\nu}^1a_1^R,\quad
\mbox{or}\ \Pi_{\mu\nu} = a_1^R.
\end{equation}

Let us calculate $a_1$ and $a_2$ using the general formula for summation over $p_4$ 
in the r.h.s.\ of Eq.\ref{15}. We may drop the terms proportional to 
$\delta_{\mu\nu}$ in r.h.s.\ of Eqs.\ref{10} and \ref{14}, because such terms vanish
after regularization and renormalization.  Then 
\begin{eqnarray}
&& \label{27} a_1 = - \frac{g^2(N^2-1)}{(2\pi)^3 2\beta} \sum_{p_4} \int d{\bf p}
\frac{(2p+k)^2}{(p^2+m^2)((p+k)^2+m^2)}, \\
&& \label{28} a_2 = - \frac{g^2(N^2-1)}{(2\pi)^3 2\beta} \sum_{p_4} \int d{\bf p}
\frac{(2p+k)_4^2}{(p^2+m^2)((p+k)^2+m^2)},
\end{eqnarray}
on account of (see [5], p.114, Eq.(7.32))
\begin{equation}\label{29}
(t^a)^{ij} (t^a)^{ji} = \frac{N^2-1}{2}.
\end{equation}
The general formula for summation over $p_4$ reads (see [7], p.123, Appendix 3, and 
[8], p.299):
\begin{eqnarray}\label{30}
{1\over\beta}\sum_n f({2\pi n\over\beta},K) = 
{1\over 2\pi}\int\limits_{-\infty}^{+\infty}
\!\!d\omega f(\omega,K)
+{1\over 2\pi}\int\limits_{-\infty+i\epsilon}^{+\infty+i\epsilon}
\!\!\!\!d\omega \frac{f(\omega,K)+f(-\omega,K)}{e^{-i\beta\omega}-1}.
\end{eqnarray}
After juxtaposition of Eqs.\ref{27},\ref{28} with Eqs.(27),(28) for statistical QED [3]
and substitution of $g^2(N^2-1)/2$ for $e^2$ we obtain corresponding expressions for
statistical quantum gluodynamics.

Eq.(36) in [3] is replaced by
\begin{eqnarray}\label{31}
\lim_{\beta\to\infty}\Pi_{\mu\nu} = 
\left(-\frac{k_\mu k_\nu}{k^2}+\delta_{\mu\nu}\right)
\left(\frac{g^2(N^2-1)}{96\pi^2}\right)k^4
\int\limits_{4m^2}^\infty\!\! dz 
\frac{\bigl(1-{4m^2}/{z^2}\bigr)^{3/2}}{z^2(z^2+k^2)}.
\end{eqnarray}
The same expression for $\Pi_{\mu\nu}$ also follows from Eq.(11) found in [9], 
where the photon Green function was calculated using the dispersion approach.

From Eq.\ref{25} and (35) from [??] we find the expression for $a_2^R$:
\begin{eqnarray}\label{32}
a_2^R = {\lambda\over3}a_1^R = 
\frac{g^2(N^2-1)k^2}{96\pi^2}(k^4-k^2)
\int\limits_{4m^2}^\infty\!\! dz 
\frac{\bigl(1-{4m^2}/{z^2}\bigr)^{3/2}}{z^2(z^2+k^2)}.
\end{eqnarray}
Finally, we get for $a_1^\beta$ and $a_2^\beta$ ($\mu\ne0$):
\begin{eqnarray}
&&\kern-10pt \label{33} a_1^\beta =
\frac{g^2(N^2-1)}{32\pi^2}
(4m^2+k^2)
\int\limits_0^\infty \frac{p\,dp}{E|{\bf k}|} 
\left(e^{\beta(E-\mu)}-1\right)^{-1} \nonumber \\ 
&&\kern-10pt\quad{}\times \ln\frac{(k^2+2p|{\bf k}|)^2+4E^2k_4^2}{(k^2-2p|{\bf 
k}|)^2+4E^2k_4^2} \\
&&\kern-10pt \label{34} a_2^\beta =
\frac{g^2(N^2-1)}{32\pi^2}
\int\limits_0^\infty \frac{p^2\,dp}{Ep|{\bf k}|} 
\left(e^{\beta(E-\mu)}-1\right)^{-1}
\biggl\{(E^2-k_4^2) \nonumber \\ 
&&\kern-10pt \quad{}\times
\ln\frac{(k^2+2p|{\bf k}|)^2+4E^2k_4^2}{(k^2-2p|{\bf k}|)^2+4E^2k_4^2} 
+ 2iEk_4
\ln\frac{(k^2+2iEk_4)^2-4p^2{\bf k}^2}{(k^2-2iEk_4)^2-4p^2{\bf k}^2}\biggr\},
\end{eqnarray}
where $E = (p^2+m^2)^{1/2}$.

\section{Conclusions}
We have proven the equivalence of many-gluon Green functions in DKP and KGF 
statistical field theories (Section~2) and calculated one-loop polarization operators
to illustrate this equivalence (Section~3).

Thus, the series of our works [1--3,9] prove that both theories lead to identical
results for {\it observable\/} physical quantities.  In this respect, it would be
interesting to prove the equivalence of the results related to the processes which 
involve unstable particles and, in particular, to apply the methods of DKP theory to 
the Standard Model.

\vskip 20pt
\noindent
{\bf Acknowledgments.} V.Ya.Fainberg is grateful to I.V.Tyutin for useful 
discussions, RFFI Fund (Grant No. 02-01-0056) and Scientific Schools Fund 
(Grant No. NS-1578.2003.2).

\end{document}